\documentclass[fleqn,twocolumn,11pt]{wlscirep}
\usepackage{gensymb}
\usepackage[utf8]{inputenc}
\usepackage[T1]{fontenc}
\usepackage[version=4]{mhchem}
\usepackage{upgreek}
\usepackage{float}
\usepackage{relsize}
\title{Water-rich amorphous state from drying mixed-metal sulfate solutions}
\author[1,*]{Christiaan T. van Campenhout}
\author[1]{Romane Le Dizès Castell}
\author[1]{Tess Heeremans}
\author[2]{Sander Woutersen}
\author[1]{Daniel Bonn}
\author[1]{Noushine Shahidzadeh}
\affil[1]{Institute of Physics, University of Amsterdam, Science Park 904, Amsterdam 1098 XH, Netherlands}
\affil[2]{Van 't Hoff Institute for Molecular Sciences, University of Amsterdam, Science Park 904,
1098XH Amsterdam, Netherlands}
\affil[*]{c.t.vancampenhout@uva.nl}

\keywords{Glasses,  Amorphous Materials, Crystallization, Sulfates, Mixed Salts}

\begin{abstract}
Amorphous and glassy materials are important for many advanced applications, from flexible solar cells to drug delivery systems. To this end, new glasses are in high demand, but precise chemical design of amorphous materials remains challenging. By studying the crystallization of mixed salt solutions, we have discovered an entirely new type of amorphous material: aqueous mixed-metal sulfate glasses. Specifically, we show that drying sulfate salt mixtures of both  mono- and higher valency cations ($2^{+} \text{or}\ 3^{+}$) almost exclusively yields a glassy state, where the stability of the amorphous state depends on the cations present and ranges from seconds to months. Furthermore, we show that the glassy state is viscoelastic, behaves like a soft solid ($G' \approx 10^{5}-10^{6}$ Pa), retains a large amount of water (30-40 w\%), and is X-ray amorphous. Additionally, confocal Raman microspectroscopy reveals disordered sulfate orientations and Fourier-transform infrared spectroscopy highlights increased hydrogen bonding during drying, which together with strong cation hydration is hypothesized to prevent crystallization. These results provide insights for the production of a new class of amorphous materials, and help to elucidate the mystery of the high abundance of such amorphous salts found on Mars.
\end{abstract}

\begin{document}

\flushbottom
\maketitle

\thispagestyle{empty}
\section*{Introduction}
Glasses and amorphous materials occupy a unique and fascinating niche in the landscape of materials science, captivating researchers not only for their fundamental roles in biomineralization processes\cite{addadi2003taking, olszta2007bone, wang2012new, stawski2016formation}, geomorphology, and planetary science\cite{rampe2020mineralogy, david2022evidence, tutolo2025carbonates}, but also as key enablers of next-generation functional materials\cite{greaves2007inorganic, wondraczek2011towards, brow2009survey}. The wide diversity of existing glasses has already unlocked a wealth of applications: amorphous solids enhance drug delivery by increasing pharmaceutical bioavailability\cite{dhaval2023review, baghel2016polymeric, pouton2006formulation, blagden2007crystal, leuner2000improving}, underpin the strength and versatility of engineering materials\cite{axinte2011glasses}, and provide innovative matrices for solid-state batteries\cite{Viallet2019,tatsumisago2008preparation}. Yet, despite their ubiquity and technological importance, the precise chemical design and predictive control of glass formation remain scientific challenges\cite{bennett2024looking}.

From this perspective, the discovery of new glassy or amorphous states is not merely an incremental advance, but a potential paradigm shift for materials innovation. In particular, complex crystallization systems, where multiple ionic species interact, offer an exciting opportunity to uncover unexpected amorphous phases. Mixed salt crystallization exemplifies this potential: for instance, while calcium carbonate typically crystallizes as calcite, the introduction of magnesium chloride can arrest this process, yielding an amorphous state and underscoring the profound influence of mixed ions on crystallization pathways\cite{raz2000formation,addadi2003taking}. However, whereas the crystallization behavior of single salts is well understood, the mechanisms and outcomes of mixed salt systems remain enigmatic and largely unexplored\cite{lindstrom2015crystallization}.

\begin{figure*}[ht]
\centering
\includegraphics[width=\linewidth]{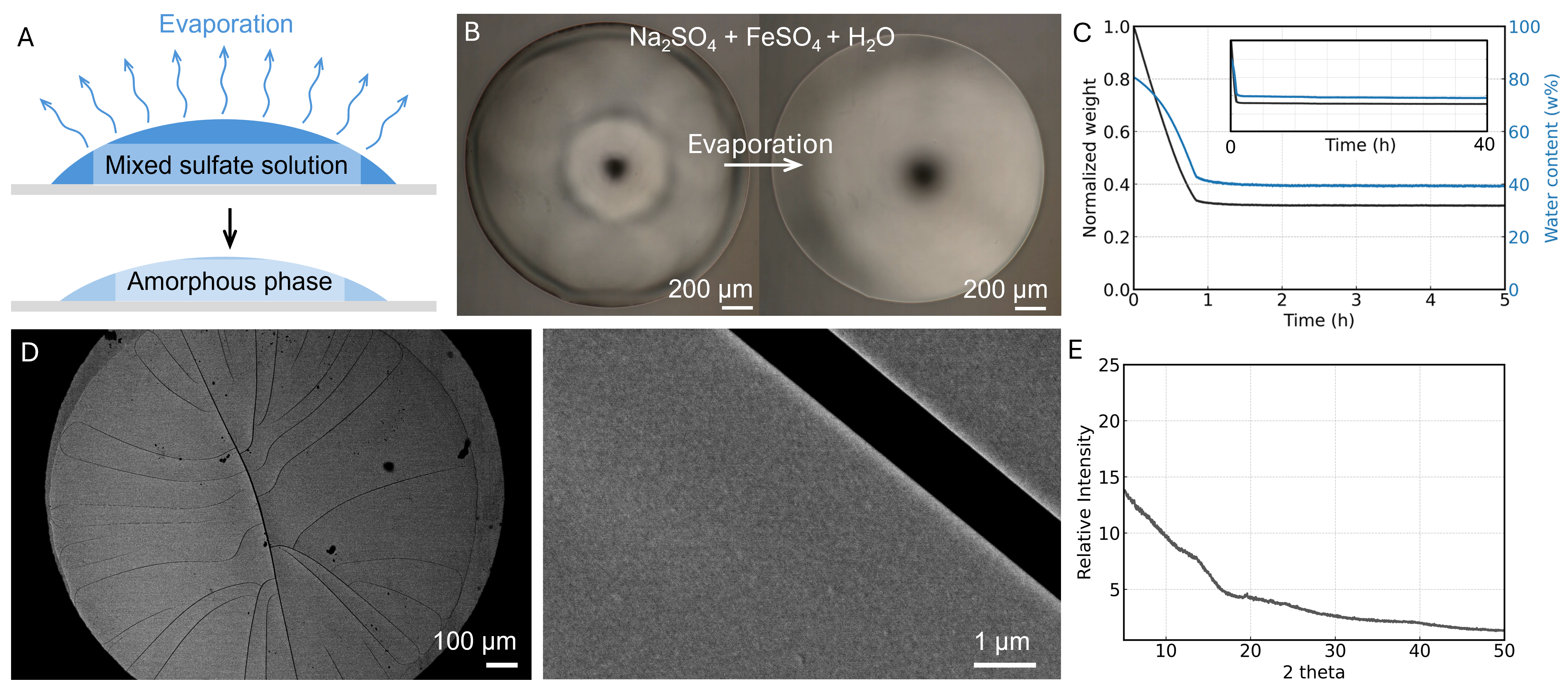}
\caption{Drying sessile droplets of mixed sulfate solutions yields an amorphous state. A) Schematic of drying of a sessile droplet. B) Phase contrast microscopy image of an amorphous state obtained through evaporation of the ternary \ce{H2O}-\ce{FeSO4}-\ce{Na2SO4} system ($V_{0} = 0.5\ \upmu\text{L}$). Evaporated at 21\degree C and RH < 10\%. See Movie S1 for a video of the droplet drying. C) Drying kinetics of a droplet ($V_{0} = 100\ \upmu\text{L},\ m_{0} = 120\ \text{mg}$) dried on a precision balance. The water and salt content are calculated from the initial weight and salt concentrations, assuming that all weight loss can be attributed to water evaporation. The inset shows the same droplet over a longer time period. D) Scanning electron microscopy (SEM) micrographs showing no internal structure on the nanoscale. Note that the cracks form in the vacuum required for sputtercoating and SEM. E) X-ray diffraction pattern showing no crystalline features.}
\label{fig:fig1}
\end{figure*}

Here, we investigate the crystallization behavior of mixed sulfate solutions and show, for the first time, the formation of aqueous mixed-metal sulfate glasses (Figure \ref{fig:fig1}A). Remarkably, we find that isothermal evaporative drying of specific mixed sulfate solutions yields a soft, viscoelastic amorphous state that retains a significant amount of water. Spectroscopic analysis indicates that a synergistic interplay between hydrogen bonding networks and strong water–cation interactions inhibits crystallization, stabilizing a metastable amorphous state. Furthermore, we demonstrate the generality of this phenomenon across a range of mixed sulfate solutions containing various mono- and multi-valent cations, where the lifetime stability of the glassy state depends on the cations present and ranges from a few seconds to months. Combined, these results open exciting new avenues for the discovery and design of functional amorphous materials.

\section*{Results and Discussion}
In order to study the formation of aqueous mixed-metal sulfate glasses, we dry sessile drops of sulfate salt mixtures through evaporation. In a typical experiment, a small droplet ($V_{0} = 0.5\ \upmu\text{L}$) of mixed sulfate solution is placed on a glass substrate and dried at room temperature, whilst tuning the evaporation rate by controlling the relative humidity. In most of this work, we study the amorphous state resulting from a 1:1 molar ratio of \ce{Na2SO4}:\ce{FeSO4} in water (Figure \ref{fig:fig1}B). Evaporative drying of such a droplet shows the formation of a glassy state, which visibly shows no further evaporation after a critical concentration is reached. To confirm this behavior, we perform a drying kinetics experiment where a larger droplet ($V_{0} = 100\ \upmu\text{L},\ m_{0} = 120\ \text{mg}$) is dried on a precision balance that tracks the weight change of the droplet over time (Figure \ref{fig:fig1}C). From this weight profile, we calculate both the salt and water content relative to the initial droplet concentrations over time. These results show that the glassy state forms when the salt concentration in the droplet has roughly tripled ([\ce{SO4^2-}]: 1.3 to 3.8 mol/kg), and that a significant amount of water (40 w\%) is trapped and remains in the droplet for days. The long-term stability of the glassy state enables a wide range of characterization techniques to study both the formation and internal structure of this material. 

To study the nanoscopic internal structure of a mixed sulfate droplet after drying, we use scanning electron microscopy (SEM). We dry a sessile aqueous droplet of \ce{Na2SO4}:\ce{FeSO4} solution on a glass substrate for 24 hours at $<10\%$ RH and sputtercoat the dried droplet with a thin gold layer (10 nm). Surprisingly, the glassy state is stable even at high vacuum required for sputtercoating and SEM analysis: we observe crack formation during vacuum pumping, but no crystallization. Interestingly, SEM analysis shows no discernible features on the nanoscale, indicating that the glassy state is indeed amorphous (Figure \ref{fig:fig1}D). To further study the crystallinity of the glassy state, we perform transmission X-ray diffraction (XRD) measurements on a droplet dried on XRD foil. Again, no crystalline features are observed, and a clear amorphous bump is present, confirming the disordered nature of the glassy state (Figure \ref{fig:fig1}E).

\begin{figure}[ht]
\centering
\includegraphics[width=\linewidth]{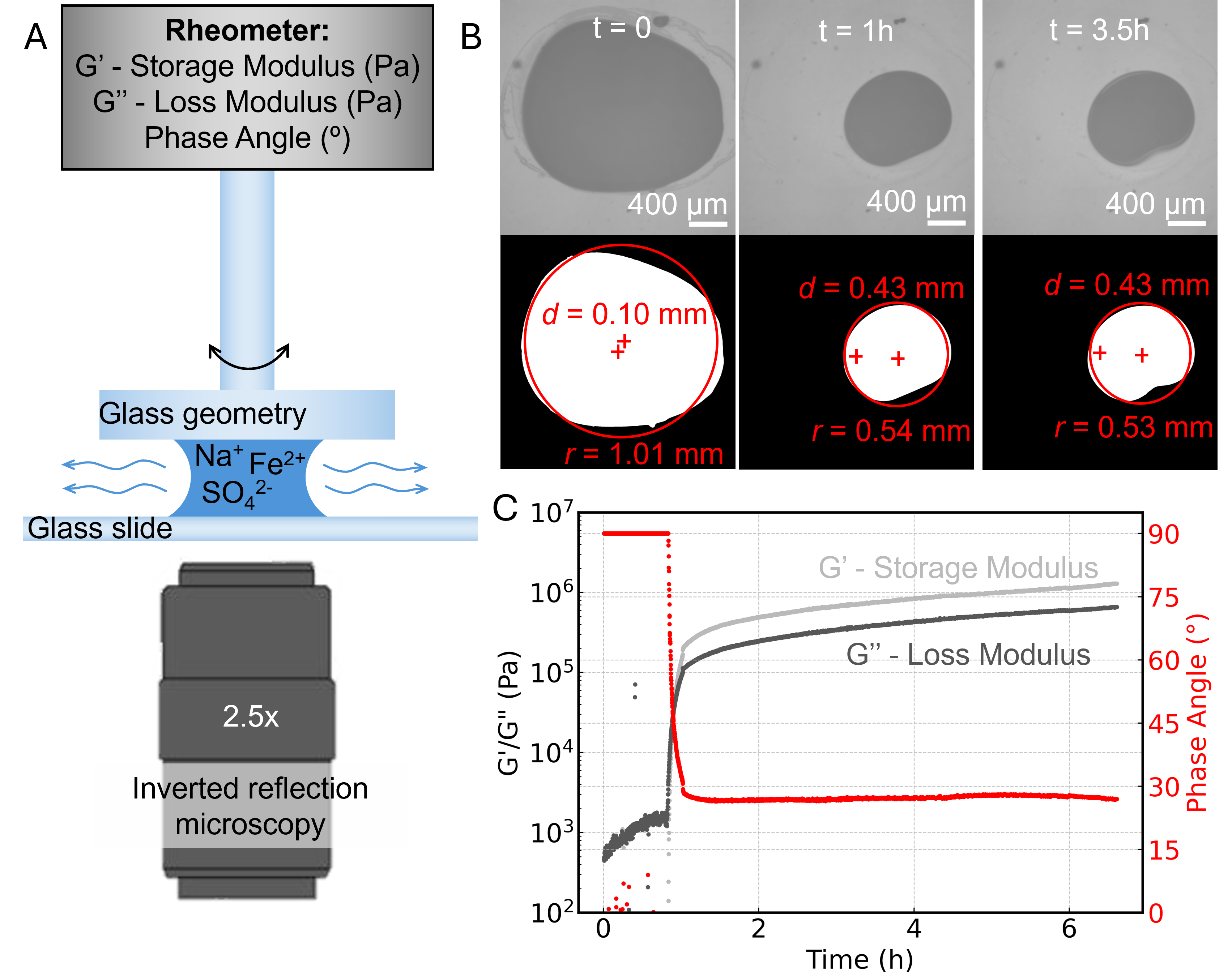}
\caption{Microscopy coupled rheology. A) Schematic of the setup used, where a rheometer head is placed on top of a reflection microscope to track changes in storage and loss moduli, whilst simultaneously monitoring the size and position of the droplet. B) Microscopy images of a droplet ($V_{0} = 0.6\ \upmu\text{L}$) containing \ce{FeSO4}-\ce{Na2SO4} in water at three time points during the measurement. The microscopy data is used to correct the rheology results for a non-centered shrinking droplet. C) Combined plot showing the storage modulus, loss modulus and phase angle changes during the evaporation of the ternary \ce{H2O}-\ce{FeSO4}-\ce{Na2SO4} system.}
\label{fig:fig2}
\end{figure}

To assess the physical properties of the glassy state, we perform rheology experiments with a custom built microscopy setup. In short, a rheometer head equipped with a glass plate-plate geometry is placed directly above an inverted reflection microscope (Figure \ref{fig:fig2}A, see Methods for additional details)\cite{desarnaud2016pressure}. This setup allows for simultaneous monitoring of visco-elastic properties of the material, as well as the shape, size and position of the droplet during evaporation. Because the droplet does not fill the entire gap of the plate-plate setup, and does not remain centered, the rheometer overestimates both the storage and the loss modulus. To correct for this overestimation, we track the radius ($r$) and distance from the droplet center to geometry center ($d$, Figure \ref{fig:fig2}B), and correct the storage and loss moduli using $G = G_{ap} \frac{4R^{4}}{6r^{2}d^{2}+3r^{4}}$, where $G_{ap}$ is the overestimated modulus reported by the rheometer and $R$ is the radius of the geometry (12.5 mm)\cite{shahidzadeh2024crystallization}. The rheology data shows a sharp increase in both storage and loss modulus at around 1 hour, at which point the size of the droplet also stops decreasing, indicating the formation of the glassy state (Figure \ref{fig:fig2}C). The size of the droplet when the glassy state is formed confirms that indeed a significant amount of water is trapped. Additionally, rheology shows that the formed glassy state acts as a soft solid, with $G' = 10^5 - 10^6$ Pa, which is comparable to silicone rubber\cite{sun2019shape}. 

\begin{figure}[ht]
\centering
\includegraphics[width=\linewidth]{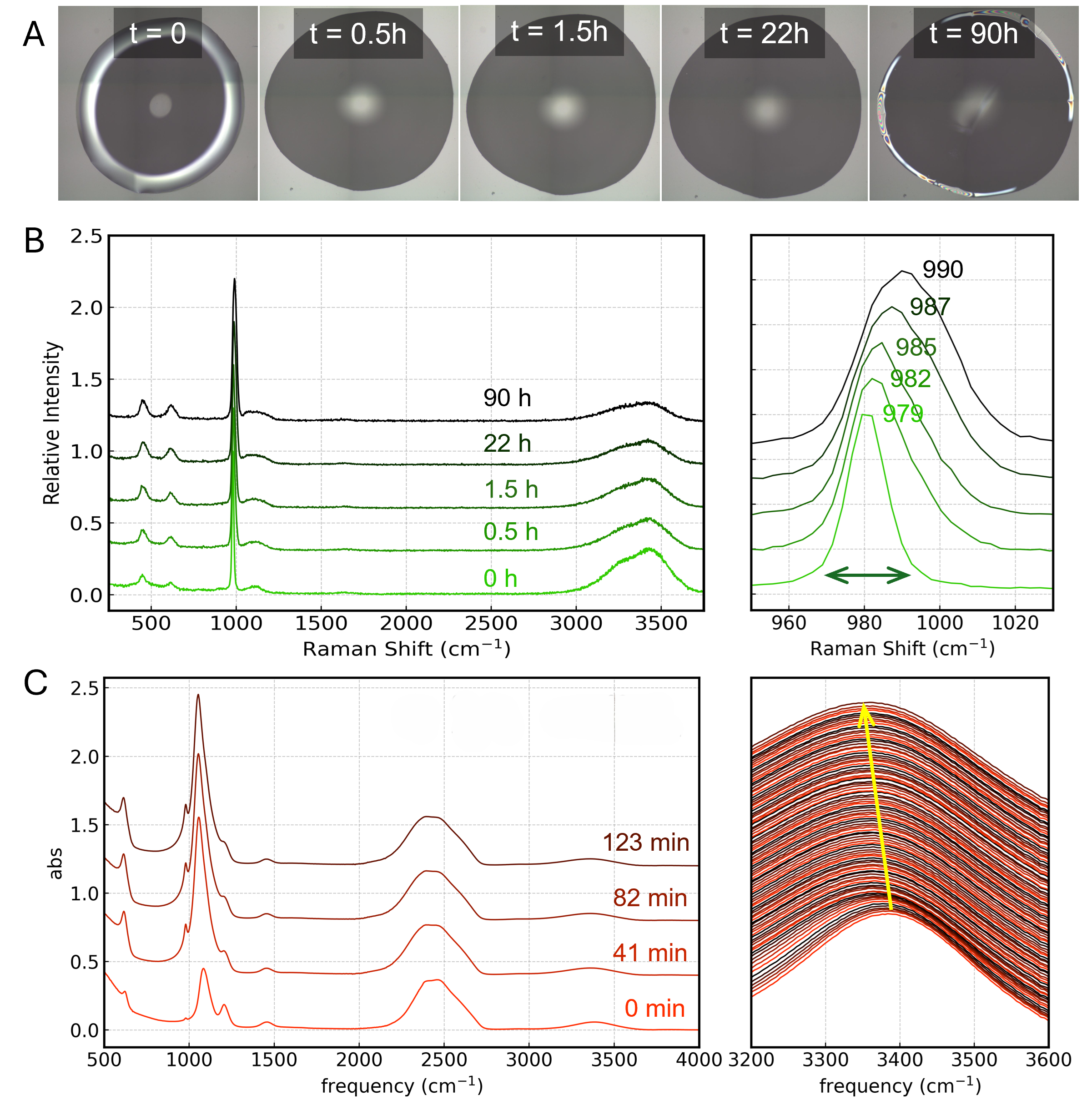}
\caption{Spectroscopic analysis of the \ce{FeSO4}-\ce{Na2SO4}-\ce{H2O} amorphous state. A) Microscopy images taken with the Raman microscope at different time points during the experiment. B) Raman spectra of an evaporating aqueous \ce{FeSO4}-\ce{Na2SO4} solution at different time points, showing a clear broadening of the $\upnu_{1}$ sulfate peak. C) Fourier-Transformed Infrared spectra of a drying \ce{FeSO4}-\ce{Na2SO4} solution (in \ce{D2O}) at different time points, showing a clear downshift of the O-H stretch in \ce{HDO}. The enlarged spectra (right) show a total time period of 2 hours.}
\label{fig:fig3}
\end{figure}

\begin{figure*}[!ht]
\centering
\includegraphics[width=\linewidth]{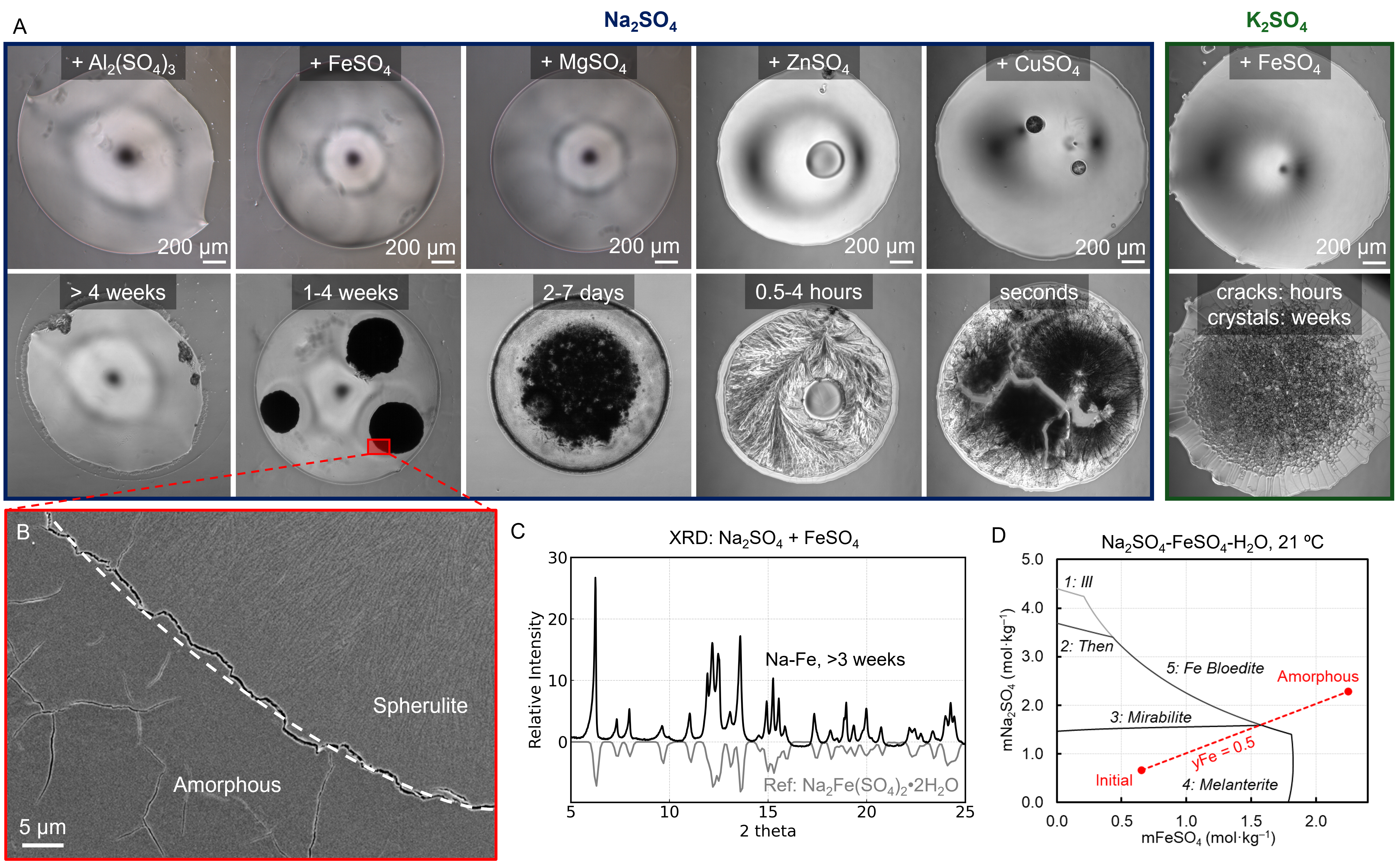}
\caption{Amorphous state observed for a wide variety of sulfate salt mixtures. A) Mixtures of sodium and potassium sulfate with various di- and trivalent cations in water, all yielding an amorphous state during drying. Based on the mixture composition, a great difference in amorphous state stability is observed, ranging from seconds to many weeks. B) SEM micrographs of an \ce{FeSO4}-\ce{Na2SO4}-\ce{H2O} droplet after 4 weeks at 30\% RH, showing the boundary of a spherulite which slowly grows over the course of weeks. C) X-ray diffraction pattern of an \ce{FeSO4}-\ce{Na2SO4}\ce{H2O} droplet after 3 weeks (after spherulite formation), showing that the metastable amorphous state decomposes into a hydrated mixed sodium iron sulfate (\ce{Na2Fe(SO4)2*2H2O}) double salt. Reference spectrum: ICSD194362\cite{barpanda2014krohnkite}. D) Equilibrium phase diagram of the ternary \ce{FeSO4}-\ce{Na2SO4}-\ce{H2O} system, showing the different equilibrium crystal phases\cite{Heeremans2025controlled}, with in red the trajectory of a droplet during drying. 1: \ce{Na2SO4} phase III, 2: \ce{Na2SO4} phase V Thenardite, 3: \ce{Na2SO4*10H2O} Mirabilite, 4: \ce{FeSO4*7H2O} Melanterite, 5: \ce{Na2Fe(SO4)2*4H2O} Iron Bloedite.}
\label{fig:fig4}
\end{figure*}

To investigate the temporal evolution of the internal chemical structure of the glassy state while drying, we apply two spectroscopic techniques. First, we investigate the structural ordering of the sulfate groups in the glassy state with confocal Raman microspectroscopy. Sulfate groups show a characteristic $\nu_{1}$ vibration at roughly $1000\ \text{cm}^{-1}$, and the shape of this peak gives information on the structural ordering of these groups.\cite{zallen1998physics} Here, we measure Raman spectra of a drying \ce{FeSO4}-\ce{Na2SO4}-\ce{H2O} droplet at different time points (Figure \ref{fig:fig3}A, B). At $t = 0\ \text{h}$ we observe the solution, and find a relatively sharp $\nu_{1}$ at $979\ \text{cm}^{-1}$. In solution, the sulfate ions are mobile and orient themselves freely, resulting in this relatively narrow peak. However, after drying ($t > 0.5\ \text{h}$), we find that this $\nu_{1}$ shifts slightly and broadens significantly, indicating that the sulfate ions are trapped in a disordered state. 

Secondly, we perform Attenuated Total Reflection Fourier Transformed Infrared Spectroscopy (ATR-FTIR) to study the water present in the glassy state. Specifically, we dry a droplet of mixed \ce{FeSO4}-\ce{Na2SO4} solution in \ce{D2O} in a nitrogen environment, and investigate the OH-stretch of \ce{HDO} (at 3300-3400 cm$^{-1}$, Figure \ref{fig:fig3}C). During evaporation, this peak significantly downshifts, indicating strengthening of the hydrogen bonding network present in the droplet. Combining Raman and FTIR results, we find that a strong hydrogen bonding network forms during drying, which contributes to preventing crystallization, resulting in a glassy state with entrapped water. 

To demonstrate the generality of this process, we screen a wide variety of different sulfate mixtures (Figure \ref{fig:fig4}A). We find that for any combination of monovalent ions (\ce{K+} or \ce{Na+}) and multivalent ions (\ce{Cu^2+}, \ce{Zn^2+}, \ce{Mg^2+}, \ce{Fe^2+}, \ce{Al^3+}) a salt ratio exists that yields an amorphous state upon drying (see Methods). Additionally, mixing only monovalent (i.e. Na-K) or only multivalent (i.e. Mg-Al) does not yield a glassy state, which suggests that the mixed ions should significantly differ in size for a stable amorphous state to form. Interestingly, the stability lifetime ($\tau_{glass}$) of the different amorphous states varies greatly, ranging from several weeks (Na-Al) to only a few seconds (Na-Cu). It is known that metal-water complexes are predominant in aqueous solutions of many metal salts, and that highly charged metal ions exchange their bound water molecules slowly with the environment. For example, the exchange rate of water bound to \ce{Al^{3+}} ions is $10^{9}$ times slower than that of water bound to \ce{Na^{+}} (Table \ref{table:cations})\cite{helm2005inorganic}.

We find that this water exchange rate, and not the ionic radius of the cations, perfectly explains the lifetime stability of the various mixed sulfate glasses: the stronger the metal-water bond of a specific cation, the more stable its respective amorphous state is. This holds true for all metals tested here, with the only outlier being \ce{Fe^2+} ($k_{H_{2}O} = 10^{-7}\ s^{-1} $), but since this readily oxidizes to \ce{Fe^3+} ($k_{H_{2}O} = 10^{-2}\ s^{-1} $) in the presence of oxygen and water, its stability is most likely attributed to a mixture of both ions. All amorphous materials are metastable, and will eventually crystallize into a more stable crystal form. Here we highlight the Na-Fe glassy state, which after several weeks shows slow growth of several spherulitical crystals (Figure \ref{fig:fig4}B). XRD analysis of these crystals reveal that a Krohnkite-type (\ce{Na2Fe(SO4)2*2H2O}) crystal phase is formed (Figure \ref{fig:fig4}C), and not the iron Bloedite (\ce{Na2Fe(SO4)2*4H2O}) phase expected from the ternary \ce{FeSO4}-\ce{Na2SO4}-\ce{H2O} phase diagram\cite{Heeremans2025controlled} (Figure \ref{fig:fig4}D).
\begin{table}
\begin{center}
\begin{tabular}{l|l|l|l}
\textbf{Cation} & $\mathbf{r_{ion}}$ \textbf{(pm)}\cite{shannon1976revised} & $\mathbf{k_{\ce{H2O}}\ (s^{-1})}$\cite{helm2005inorganic}& $\mathbf{\tau_{glass}}$
\\ \hline
\ce{Al^3+}  & 54   & $10^{0}$ & >4 weeks         \\
\ce{Fe^3+}  & 65     & $10^{2}-10^{3}$ & n.a.           \\
\ce{Fe^2+}  & 78     & $10^{6}-10^{7}$ & 1-4 weeks           \\
\ce{Mg^2+}  & 86     & $10^{5}-10^{6}$ & 2-7 days            \\
\ce{Zn^2+}  & 74     & $10^{7}-10^{8}$ & <4 hours         \\
\ce{Cu^2+}  & 73     & $10^{9}-10^{10}$ & seconds       \\ 
\ce{Na+}    & 112    & $10^{8}-10^{9}$ & n.a.        \\ 
\ce{K+}     & 146    & $10^{8}-10^{9}$ & n.a.     \\\hline
\end{tabular}
\caption{Ionic radii ($r_{ion}$), water exchange rates ($k_{\ce{H2O}}$) and glassy state lifetime ($\tau_{glass}$) of the cations used in this work. The lifetime is given for the cation-\ce{Na+} mixed salt.}
\label{table:cations}
\end{center}
\end{table}
\section*{Conclusion}
In summary, we introduce here an unexpected new class of amorphous materials: aqueous mixed-metal sulfate glasses. Through evaporative drying of mixed sulfate solutions, we create a wide variety of soft, viscoelastic and water containing glassy states. We find that a strong hydrogen bonding network, in combination with metal-water complexes immobilizes all species inside the droplet, inhibiting crystallization and thus yielding a metastable amorphous state. Furthermore, we show that a glassy state only forms for mixtures of mono- and multi-valent cations. 

Our results highlight the complexity of mixed salt solutions, but further research is required to fully understand the mechanism underlying the formation of these glassy sulfates. We envision that a combination of computational techniques and high resolution electron microscopy can provide fundamental insights on the molecular scale.

Besides this fundamental interest, we believe that these amorphous sulfates enable a wide range of follow-up research. For instance, the materials discussed in this work, and specifically iron sodium mixed sulfates, closely resemble the materials investigated for solid state battery technology\cite{zhang2023bridging, chakraborty2018current}, and their facile synthesis could enable research towards advanced energy system applications. Additionally, our results provide important insights for the fields of biomineralization and planetary science, where amorphous states are often observed and crystallization processes almost exclusively involve mixed solutions. Specifically, we foresee that our findings will provide key clues into the mystery of the high abundance of amorphous sulfate materials found on Mars, but not on Earth\cite{rampe2020mineralogy, david2022evidence}.
\smaller[1]
\section*{Methods}
\subsection*{Mixed sulfate solution preparation}

All chemicals were used without additional purification and purchased from Sigma-Aldrich, with $>99\%$ purity. Solutions were prepared directly prior to use following this general procedure: to \ce{H2O} was added \ce{Na2SO4} and \ce{FeSO4*7H2O}. The solution was thoroughly vortexed and sonicated to ensure complete dissolution. Note that the ratio of cations used is crucial for the formation of an amorphous state, and all ratios given here yield the most stable amorphous state.

\begin{table}[H]
\smaller[1]
\begin{center}
\begin{tabular}{l|l|l}
\textbf{Mix} & \textbf{Salt 1} & \textbf{Salt 2} 
\\ \hline
\ce{Na+}:\ce{Al^3+}, 4:1 & \ce{Na2SO4}   & \ce{Al2(SO4)3*18H2O}          \\
\ce{Na+}:\ce{Fe^2+}, 2:1 & \ce{Na2SO4}   & \ce{FeSO4*7H2O}             \\
\ce{Na+}:\ce{Mg^2+}, 4:1 & \ce{Na2SO4} & \ce{MgSO4*7H2O}            \\
\ce{Na+}:\ce{Zn^2+}, 5:1 & \ce{Na2SO4} & \ce{ZnSO4*7H2O}           \\
\ce{Na+}:\ce{Cu^2+}, 5:1 & \ce{Na2SO4}   & \ce{CuSO4*5H2O}           \\
\ce{K+}:\ce{Fe^2+}, 1:1  & \ce{K2SO4}    & \ce{FeSO4*7H2O}        \\ \hline
\end{tabular}
\caption{Salts used for amorphous states.}
\label{table:table_materials}
\end{center}
\end{table}

\subsection*{Spectroscopy and Electron Microscopy}
Raman microspectroscopy was performed using a WITec Alpha 300R microscope coupled to a CMOS camera (Andor, Newton EMCCD, DU970P-BVF-355). The wavelength used was 532 nm, with a $600\ mm^{-1}$ diffraction grating. Raman droplet $V_{0} = 0.7\ \upmu\text{L}$. Scanning electron microscopy was performed using a FEI Verios 460 scanning electron microscope. SEM droplet $V_{0} = 0.7\ \upmu\text{L}$. ATR-FTIR spectra were measured on a Perkin-Elmer Frontier FT-IR spectrometer fitted with a Pike GladiATR module that included a heated top plate and a diamond ATR-crystal. A droplet $V_{0} = 2\ \upmu\text{L}$ of mixed sulfate solution in \ce{D2O} was placed on the diamond ATR-crystal and a constant nitrogen flow was applied to prevent exchange of \ce{D2O} with \ce{H2O} from the surroundings.

\subsection*{XRD}
The XRD measurements were performed on a STADI P diffractometer (STOE\&Cie) with Ge-monochromated $\ce{Mo}-\ce{K}\alpha$1 radiation ($\lambda = 71.073$ pm) in Debye-Scherrer geometry. Samples were prepared by drying a mixed sulfate solution droplet ($V_{0} = 20\ \upmu \text{L}$) on XRD foil at room temperature. Before measuring, a second XRD foil was placed over the droplet, effectively sandwiching the droplet between two foils. Care was taken not to exert excessive pressure on the droplet, as this can induce crystallization. 

\subsection*{Microscopy Coupled Rheology}
Rheology experiments were performed using a custom built reflection microscopy coupled rheology setup, based on the setup used in previous work\cite{desarnaud2016pressure}. An Anton Paar DSR502 Measuring Head is placed on top of a reflection microscope using a custom built mount. A custom glass plate-plate geometry is used, and a $V_{0} = 0.6\ \upmu\text{L}$ droplet is placed on a glass slide. The geometry is then brought to the measuring distance of 100 $\upmu$m, where it makes good contact with the droplet. The measurement is an oscillatory test, with an amplitude $\gamma = 5\%$ and frequency $f = 1\ \text{Hz}$, which gives both the storage and loss moduli, as well as the phase angle. Simultaneously, the reflection microscope records images and a custom Python3 script using Canny edge detection\cite{canny1986computational} finds the position and size of the drying droplet.
\bibliography{main}

\section*{Acknowledgements}
We thank Dr. Simon Lepinay and Rowen Hersche for their work on preliminary experiments, and Dr. Falk Pabst for his help with XRD measurements. 

\section*{Author contributions statement}
C.C., R.C., S.W., D.B. and N.S. designed and performed experiments. T.H. performed preliminary experiments. C.C., S.W., D.B. and N.S. wrote and edited the manuscript.

\section*{Additional information}

\textbf{Competing interests} The authors declare no competing interest. 
\end{document}